\documentstyle[12pt]{article}

\newcommand{\be}{\begin{equation}}
\newcommand{\ee}{\end{equation}}
\newcommand{\bea}{\begin{eqnarray}}
\newcommand{\eea}{\end{eqnarray}}

\newcommand{\newsection}[1]{
\vspace{10mm}
\pagebreak[3]
\addtocounter{section}{1}
\setcounter{footnote}{0}
\begin{flushleft}
{ \bf{ \thesection. #1}}
\end{flushleft}
\nopagebreak
\vspace{4mm}
\nopagebreak}

\newcommand{\NP}[1]{Nucl. Phys.\ {\bf #1}\ }
\newcommand{\PL}[1]{Phys. Lett.\ {\bf #1}\ }

\newcommand{\PR}[1]{Phys. Rev.\ {\bf #1}\ }
\newcommand{\PRL}[1]{Phys. Rev. Lett.\ {\bf #1}\ }

\newcommand{\MPL}[1]{Mod. Phys. Lett.\ {\bf #1}\ }

\newcommand{\CR}{\nonumber \\}

\newcommand{\LM}{\Lambda}
\newcommand{\vp}{\varphi}

\begin{document}
\begin{titlepage}
\begin{flushright}
{\large \bf UCL-IPT-95-22}
\end{flushright}
\begin{center}
\vskip 2cm

{\Large \bf
Is the standard Higgs scalar elementary?}
\vskip .4in

{\large D. Delepine\footnote{Research assistant of the National Fund for
the Scientific Research}, J.-M. G\'erard and R. Gonzalez Felipe}\\[.15in]

{\em Institut de Physique Th\'eorique\\
 Universit\'e catholique de Louvain\\
B-1348 Louvain-la-Neuve, Belgium}

\end{center}

\vskip 2in

\begin{abstract}

In the standard electroweak model, the measured top quark mass requires a
sizeable Yukawa coupling to the fundamental scalar. This large
coupling alone might induce a dynamical breaking of the electroweak symmetry
 as well as non-perturbative effects. If such
is the case, even a standard Higgs scalar  as light as 80~GeV should have a
non-negligible $t \bar{t}$ component induced by the top condensate.
\end{abstract}

\end{titlepage}

\newpage
\renewcommand{\thepage}{\arabic{page}}
\setcounter{page}{1}
\setcounter{footnote}{1}

\newsection{Introduction.}

Well before the advent of QCD, Nambu and Jona-Lasinio (NJL) \cite{NJL}
introduced a four-fermion interaction to break the chiral symmetry of
strong interactions. In modern language, this NJL effective Lagrangian is
expected to be induced by multiple gluon exchanges in the light
quark-antiquark channels.

The electroweak gauge-couplings of the Standard Model are not strong
enough to trigger a similar breaking of the flavour $SU(2)_L \times U(1)$
 gauge symmetry. But new gauge interactions beyond the Standard Model
might generate effective four-fermion interactions
\cite{nambu}-\cite{BHL}. In that approach, the scalar field $h$
responsible for the symmetry breaking is a pure
$t \bar{t}$ composite state with $m_h = 2 m_t$, if QCD effects
are ignored.

It is however quite remarkable that the strongest force in the
electroweak sector of the Standard Model is due to the Yukawa coupling of
the recently observed top quark \cite{CDF} to the fundamental
Higgs field . It is therefore
quite legitimate to investigate the possibility of a $t\bar{t}$
condensation without having to invoke new physics beyond the Standard
Model \cite{ross}. Indeed, the Yukawa coupling {\it itself} might generate
a four-fermion interaction at some scale $\LM$, in a way similar
to what is happening in QCD.

In this letter, we analyze the implications of this minimal scenario on
 the scalar mass spectrum. For that purpose, we assume that
the $SU(2)_L \times U(1)$ gauge couplings and the scalar self-coupling
do not participate at all in the symmetry breaking. In that case, the
electroweak symmetry breaking is also triggered by top quark loops
\cite{fatelo} such that the standard Higgs boson is a linear
combination of the $t\bar{t}$ composite state and of the fundamental
scalar. In particular, if the $\LM$ scale  is around 1 TeV, the
elusive Higgs scalar is mainly a composite state with a mass of about
80 GeV.

\newsection{The scalar Lagrangian.}

Let us assume that the Yukawa coupling $g_t$ of the top quark alone is indeed
responsible for the electroweak symmetry breaking below the cutoff scale
$\LM$. If such is the case, the relevant Lagrangian
for the fundamental iso-doublet scalar field $H$  simply becomes

\be
L_H= \partial_\mu H^+ \partial^\mu H - m_H^2 H^+ H +
g_t (\bar{\psi}_L t_R H + h.c.)\ ,
\ee
where $H=\left(\begin{array}{l}H^0\\H^- \end{array}\right)$
and $\psi_L=\left(\begin{array}{l}t_L\\b_L \end{array}\right)$.
\\

The crucial non-perturbative effect possibly induced by a large $g_t$ Yukawa
coupling  is the appearance of a 4-fermion interaction

\be
L_{NJL} = G \bar{\psi}_Lt_R\bar{t}_R\psi_L\ ,
\ee
after resummation of the multiple scalar
exchanges in the heavy quark-antiquark channels.
The 4-fermion form-factor
is expected \cite{ross} to depend on the  $q^2$ transfer-momentum  in the
following way

\be
G(q^2)=\frac{g_\sigma^2}{q^2-m_\sigma^2}\ ,
\ee
\\
\noindent
with $g_\sigma$, an effective coupling, and $m_\sigma$, an effective mass.

Implications of this possible non-perturbative effect are most easily
described in terms of the iso-doublet auxiliary field

\be
\Sigma \equiv \frac{g_\sigma}{q^2-m_\sigma^2} \bar{t}_R \psi_L\ ,
\ee
such that the Lagrangian

\be
L_{NJL} = \partial_\mu \Sigma^+ \partial^\mu \Sigma - m_\sigma^2 \Sigma^+
\Sigma + g_\sigma (\bar{\psi}_L t_R \Sigma + h.c.)\ .
\ee
\\
\noindent
is equivalent to Eq.(2).
The scalar Lagrangian $L_H + L_{NJL}$ at the source of the $SU(2)_L \times
U(1)$ breaking below the scale $\LM$ is then defined by Eqs.(1) and (5).

Notice that the absence of a kinetic term for the $\Sigma$ field at the
scale $\LM$ would imply the compositeness boundary condition \cite{BHL}
$g_\sigma(\LM)=\infty$ on the physically normalized
coupling $g_\sigma$. But  the coupled
renormalization group equations for $g_t$ and $g_\sigma$ require
the ratio  $g_\sigma/g_t$ to be independent of the scale. This would
 obviously be in
contradiction with the Lagrangian given in Eq.(1) where
$g_t(\mu)$ is supposed to have a fixed and finite value at
the $\LM$ scale.

This kinetic term is important since it allows the $\Sigma$ field to
contribute also to the $W$ gauge-boson mass through the covariant derivative,
once the (small) $SU(2)$
gauge coupling $g_2$ is switched on. Therefore, both the loop-induced
vacuum expectation values  of the elementary field $H^0$
{\it and} of the composite field $\Sigma^0$ contribute to the $W$ gauge-boson
mass

\be
m_W^2=\frac{1}{2} g_2^2 (\langle H^0 \rangle^2+\langle \Sigma^0
\rangle^2)
\equiv \frac{1}{4} g_2^2 (\langle \vp \rangle^2+\langle \sigma
\rangle^2)\ ,
\ee

\noindent
and to the top quark mass

\be
m_t=g_t \langle H^0 \rangle + g_\sigma \langle \Sigma^0 \rangle
\equiv \frac{1}{\sqrt{2}}( g_t \langle \vp \rangle + g_\sigma \langle
\sigma \rangle) \ .
\ee

\newsection{The top-induced effective potential.}

Now, we shall study how the electroweak symmetry is fully induced by top
quark loops \cite{fatelo}. If we focus on the real part of the neutral $H^0$
and $\Sigma^0$ components in the scalar Lagrangian $L_H+L_{NJL}$, the
one-loop effective potential reads

\be
V(\vp,\sigma)=m_H^2 \frac{\vp^2}{2}+m_\sigma^2 \frac{\sigma^2}{2}
-\frac{N_c}{8\pi^2} \int^{\LM^2}_{0}\ q^2 \ln \left( 1+ \frac{m_t^2}{q^2}
\right) dq^2 \ ,
\ee
\\
\noindent
with $N_c$, the number of colours.

 The extrema conditions are given by

\bea
\left. \frac{\partial V}{\partial \vp}\right|_{\langle \vp \rangle,
\langle \sigma \rangle} &=& m_H^2 \langle \vp \rangle
-\frac{N_c}{8\pi^2} \sqrt{2} g_t m_t \int^{\LM^2}_{0}\ \frac{q^2}{q^2+m_t^2}
dq^2 =0\ , \CR
\left. \frac{\partial V}{\partial \sigma}\right|_{\langle \vp \rangle,
\langle \sigma \rangle}&=& m_\sigma^2 \langle \sigma \rangle
-\frac{N_c}{8\pi^2} \sqrt{2} g_\sigma m_t \int^{\LM^2}_{0}\
\frac{q^2}{q^2+m_t^2} dq^2 =0 \ .
\eea
\\
\noindent
{}From Eqs.(7) and (9), we obtain then the self-consistent relation

\be
\frac{m_H^2 m_\sigma^2}{(g_\sigma^2m_H^2+g_t^2m_\sigma^2)} =
\frac{N_c}{8\pi^2} \left(\LM^2-m_t^2\ln(\frac{\LM^2}{m_t^2}+1)\right)
\equiv I_1
\ee
\\
\noindent
and the vacuum expectation values

\bea
\langle \vp \rangle &=& \sqrt{2} g_t \frac{m_t}{m_H^2} I_1 \ ,\CR
\langle \sigma \rangle &=& \sqrt{2} g_\sigma \frac{m_t}{m_\sigma^2} I_1 \ .
\eea
\noindent
On the other hand, Eq.(6) requires the following normalization

\be
v=\sqrt{\langle \vp \rangle^2+\langle \sigma \rangle^2}\ =
 246\ {\rm GeV}\ ,
\ee
for these vacuum expectation values.

As it should be (see Eq.(3)), in the limit $m_\sigma^2 \rightarrow \infty,
\langle \sigma \rangle \rightarrow 0$ and we recover the model considered
in Ref.\cite{fatelo}, with only one elementary scalar iso-doublet.

The gap equation in (10) can also be derived by requiring the
existence of a Goldstone boson in the pseudoscalar neutral sector. This
constraint is indeed fulfilled  if the determinant of the neutral pseudoscalar
squared mass matrix $M_{PS}^2$

\be
M_{PS}^2=\left(
\begin{array}{cc}
 m_H^2-g_t^2I_1	 & -g_tg_\sigma I_1	 \\
 -g_tg_\sigma I_1	 & m_\sigma^2-g_\sigma^2I_1
\end{array}
\right)
\ee
\\
\noindent
is vanishing. The physical basis for the neutral pseudoscalars is
obtained  from  the diagonalization of $M_{PS}^2$ through a rotation of angle
$\theta_{PS}$ with

\be
\tan \theta_{PS} = \frac{\langle \sigma \rangle}{\langle \vp \rangle}\ .
\ee

\newsection{The standard Higgs scalar.}

The squared mass matrix $M_S^2$ for the neutral $\vp$ and $\sigma$ scalar
fields is obtained after the substitution $I_1 \rightarrow I_1-I_2$ with

\bea
I_2 &=& \frac{N_c}{4\pi^2} m_t^2 \int_0^{\LM^2} \frac{q^2 dq^2}
{(q^2+m_t^2)^2} \CR
&=& \frac{N_c}{4\pi^2} m_t^2 \left(\ln(\frac{\LM^2}{m_t^2}+1)
-\frac{\LM^2}{\LM^2+m_t^2}\right)\ ,
\eea
in the pseudoscalar squared mass matrix given in Eq.(13). Consequently, we
now have to diagonalize the $2 \times 2$ matrix

\be
M_S^2=M_{PS}^2+\left(\begin{array}{cc} g_t^2 & g_t g_\sigma\\
g_t g_\sigma & g_\sigma^2 \end{array} \right) I_2\ .
\ee
\\

As $\LM^2 \gg m_t^2$, $I_2$ is small compared to $I_1$ and the
diagonalization angle $\theta_S$ for the $M_S^2$ matrix is very close to
$\theta_{PS}$. The lightest neutral scalar field $h$ is therefore almost
in alignement with the pseudoscalar Goldstone boson:

\be
h \simeq \cos \theta_{PS}\ \vp + \sin \theta_{PS}\ \sigma\ .
\ee
Its squared mass proportional to $I_2$ (see Eq.(16)) has a smooth
logarithmic dependence on the scale $\LM$ and is approximately given by

\be
m_h^2 \approx (g_t \cos \theta_{PS}+ g_\sigma \sin \theta_{PS})^2 I_2\ .
\ee
The other scalar field has a mass proportional to the scale $\LM$.

{}From Eqs.(7), (14) and (18), we conclude that the mass $m_h$ of the
standard Higgs scalar $h$ is equal to

\be
m_h \approx \left( \frac{2 I_2}{v^2} \right)^{1/2} m_t
\ee
\\
\noindent
and does not depend on its structure. In particular, it can be as small
as  80~GeV if the $\LM$ scale is about 1~TeV.

The interesting mass relation given in Eq.(19) is based on the assumption
that the heavy top quark alone triggers the full electroweak symmetry
breaking. This relation has been derived in Ref.\cite{fatelo} for the
special case of a pure elementary scalar field ($\theta_{PS}=0$). However,
Eq.(17) shows that the mass relation in Eq.(19) remains valid even if the
standard Higgs scalar has a large $t \bar{t}$ component ($\theta_{PS} >
\pi/4$). Such an intriguing possibility is in fact favoured if some
$t \bar{t}$ condensation takes place at the electroweak scale.

In the Standard Model without $t \bar{t}$ condensation ($\langle \sigma
\rangle=0$), $g_t \approx 1$ corresponds to $m_t \approx 174$ GeV. But the
non-perturbative condensation mechanism ($\langle \sigma \rangle \neq 0$)
assumed in this letter requires a larger Yukawa coupling

\be
g_t \gg 1\ ,
\ee
at the electroweak scale.
This physical constraint together with the fact that $g_t \langle \vp
\rangle$  and $g_\sigma \langle \sigma \rangle$ are positive (see Eqs.(11))
 imply therefore

\be
\langle \vp \rangle \ll v
\ee
to reproduce the measured top quark mass defined in Eq.(7). From Eqs.(12)
and (21), we then obtain the following hierarchy

\be
\langle \sigma \rangle \gg \langle \vp \rangle
\ee
among the vacuum expectation values, such that the standard Higgs scalar
$h$ defined in Eq.(17) has a dominant ($\theta_{PS} > \pi/4$) $t \bar{t}$
component.

For illustration, let us assume the quite reasonable value

\be
\alpha_t \equiv \frac{g_t^2}{4 \pi} \simeq 1\ ,
\ee
\\
\noindent
for the genuine Yukawa coupling of the top quark. For $m_t=174$ GeV, we
obtain

\be
\sin \theta_{PS} > 0.96
\ee
\\
\noindent
and the standard Higgs scalar is indeed an almost pure $t \bar{t}$ bound
state.

Let us remark that for very high values of the scale $\LM$, the running
of couplings down to the electroweak scale must be taken into account. If
these couplings enter the perturbative regime above the electroweak
scale, they will approach the quasi-fixed point \cite{hill} given in our
case by

\be
g_t^2+g_\sigma^2 \approx \frac{16}{9} g_3^2\ ,
\ee
\\
\noindent
with $g_3$, the strong interaction coupling. The existence of such a point
usually leads to a too heavy top quark in top condensate models \cite{BHL}.
However, in the present case, the top quark mass is a linear combination
of the $g_t$ and $g_\sigma$ couplings and its experimental value \cite{CDF}
can be  reproduced even if the quasi-fixed point is reached.

We emphasize however that, in the approach considered here, there is no reason
whatsoever to have a very high scale $\LM$ to reproduce the $W$ gauge-boson
mass (see Eq.(6)).

\newsection{Conclusion}

Assuming that a strong Yukawa coupling of the top quark is fully responsible
for the electroweak symmetry breaking \cite{ross,fatelo}, we have shown
that a $t \bar{t}$ condensation would imply a large composite component for
a rather light standard Higgs scalar. The effective approach presented
in this letter illustrates how a dynamical symmetry breaking might, in
principle,  avoid two generic problems associated with top condensate models
\cite{nambu}-\cite{BHL}, namely a (too) heavy top quark due to
compositeness boundary conditions and a (too) high $\LM$ scale due to
loop-induced gauge-boson masses. Here, the genuine top Yukawa coupling
must be finite at the $\LM$ scale and the gauge-boson masses are induced
by $SU(2) \times U(1)$ covariant derivatives. It is also remarkable that
the mass prediction of Ref.\cite{fatelo} for the Higgs scalar remains
valid in the presence of a top condensate. In particular, the usual
relation $m_H \approx 2 m_t$ does not apply and a very light Higgs scalar
is predicted if $\LM$ is around 1 TeV.

\vspace*{1cm}
\noindent
{\large\bf Acknowledgements}\\

We are grateful to Jacques Weyers for many useful discussions and comments.
This work was supported in part by the EEC Science Project SC1-CT91-0729.


\newpage

\begin{thebibliography}{99}



\bibitem{NJL} Y. Nambu and G. Jona-Lasinio, \PR{122} (1961) 345.

\bibitem{nambu}Y. Nambu, in {\it New Theories in Physics}, proceedings of
the XI International Symposium on Elementary Particle Physics, edited by
Z. Ajduk, S. Pokorski and A. Trautman (World Scientific, Singapore, 1989).

\bibitem{miransky} A. Miransky, M. Tanabashi and K. Yamawaki, \MPL{A 4}
(1989) 1043; \PL{B 221} (1989) 177.

\bibitem{BHL} W.A. Bardeen, C.T. Hill and M. Lindner, \PR{D 41} (1990) 1647;

\bibitem{CDF} F. Abe et al. (CDF collaboration), \PRL{74} (1995) 2626;
S. Abachi et al. (D0 collaboration), \PRL{74} (1995) 2632.

\bibitem{ross} D.E. Clague and G.G. Ross, \NP{B364} (1991) 43.

\bibitem{fatelo} J.P. Fatelo, J.-M. G\'{e}rard, T. Hambye and J. Weyers,
\PRL{74} (1995) 492.

\bibitem{hill} B. Pendleton and G.G. Ross, \PL{B 98} (1981) 291;
C.T. Hill, \PR{D 24} (1981) 691.

\end{thebibliography}
\end{document}